\documentclass[reprint,amsmath,amssymb,aps,pra]{revtex4-1}
\usepackage{amsmath,amssymb,mathtools,braket}
\usepackage{graphicx}
\usepackage{dcolumn}
\usepackage{bm}
\usepackage[colorlinks=true,allcolors=blue]{hyperref}
\usepackage{float}

\begin{document}


\title{Perfectly matched layer method for optical modes in dielectric cavities}

\author{Tianpeng Jiang} \email{tjiangad@connect.ust.hk}
\author{Yang Xiang} \email{maxiang@ust.hk}
\affiliation{Department of Mathematics, The Hong Kong University of Science and Technology, Clear Water Bay, Kowloon, Hong Kong SAR, China}
\date{\today}
\begin{abstract}
The optical resonance problem is similar to but different from time-steady Schr\"{o}dinger equation in the point that eigenfunctions in resonance problems are exponentially growing.  
We introduce perfectly matched layer method and the complex stretching technique to transform eigenfunctions from exponential-growth to exponential-decay.
Accordingly, we construct a Hamiltonian operator to calculate eigenstates of optical resonance systems. 
We successfully apply our method to calculate the eigenvalues for whispering-gallery modes and the results perfectly agree with existing theory that is developed only for regularly-shaped cavities. 
We also apply the method to investigate the modes evolution near exceptional points\textemdash a special phenomenon that only happens in non-Hermitian systems.
The presenting method is applicable to optical resonance systems with arbitrary dielectric distributions.
\end{abstract}
\maketitle

\section{Introduction}
With the well-developed fabrication technology, the optical resonance phenomena in dielectric microcavities \cite{phototech} have been applied in varieties of emerging photonic technologies, such as microlasers \cite{disklasers,disklasers2010}, optical filters \cite{diskfilters}, photonic circuits \cite{photoIC2004,photoIC2014}, nanoparticle sensors \cite{sensors2011,sensors2012}, and optical gyroscopes \cite{gyro} etc. 
The resonance states formed in cavities are intrinsically lossy because optical cavities are open systems with electromagnetic energy radiating to infinity.
The openness makes the effective Hamiltonian of optical resonance systems being non-Hermitian \cite{nonhermitian2015,nonhermitian2007}, and therefore novel phenomena happen, such as wave chaos \cite{wavechaos1997,wavechaos2002,wavechaos2010} and exceptional points \cite{EPs2014,EPs2018}. 

The optical resonance problems can be analytically solved for regularly-shaped cavities such as circular cavities \cite{WGMs1910,WGMs2002}, square cavities \cite{squarecavity2001}, and rectangular cavities \cite{squarecavity2016}.
For deformed cavity shapes, analytical approximations are obtained by means of perturbation theories.
One perturbation approach is based on a perturbation ansatz for only symmetric cavities \cite{perturbation2008}, and later the ansatz is modified for asymmetric cavities \cite{perturbation2016}.
This approach is applied to calculate optical modes in cavity shapes of cut-disk \cite{perturbation2008}, Lima{\c{c}}on \cite{limacon2014}, spiral \cite{perturbation2016} and polar-cosine \cite{polar}.
Another perturbation approach is based on rigorous perturbation theory without presumed ansatz and the method is justified as successfully applied to Lima{\c{c}}on and spiral cavities \cite{perturbation2019}.
For largely deformed and more general cavity shapes, numerical solutions are necessary. A boundary-element based numerical method has been developed, but spurious solutions exist in that method and the issue has been discussed \cite{BEM}.

Perfectly matched layer (PML) is an artificial absorbing layer at far field region for solving acoustic and electromagnetic wave equations.
The PML method is first proposed by B\'{e}renger for wave scattering problems \cite{PMLBerenger1994,PMLBerenger1996}, and the original formulation involves field splitting in the absorbing layer.
Then Chew and Weedon \cite{PMLChew1994}, avoiding this splitting, realize that B\'{e}renger's formulation is equivalent to a complex-coordinate stretching. 
Based on the complex stretching, PML equations for curvilinear coordinates \cite{PML199801,PML199802,PML2005} and for convex geometries \cite{PML2001} are developed, with analysis on the existence and uniqueness of PML solutions proved \cite{PMLexistence1998,PML2009}.
The PML methods has also been adopted in solving resonance problems in open systems for fluid dynamics \cite{PMLfluid}, aero acoustics \cite{PMLacoustic} and electromagnetics \cite{PML2007}.

In this paper, based on PML we develop a novel method to calculate optical modes in cavities with arbitrary dielectric distributions.
Our method does not impose any requirements on the cavity shapes, so it is applicable to optical resonance systems including largely-deformed cavities, multi-cavities and random media etc. 
Additionally, with the complex stretching technique, the eigenfunctions of resonance problems are transformed from exponential-growth to exponential-decay, and hence the eigenfunctions are restricted in a certain Hilbert space, which can help to formulate the optical resonance problems under the framework of quantum mechanics.

We successfully validate our method by applying to circular cavities calculating the whispering-gallery modes (WGMs), the results agree well with existing theory which was developed only for regularly-shaped cavities.  
Because the dielectric cavities are open systems with energy radiating to infinity, the resonance systems and the proposed Hamiltonian operator are non-Hermitian.
To validate that our method reflects the non-Hermitian characteristics, we apply the method to quad-cosine cavities to investigate modes evolution near exceptional points (EPs), which is a special phenomenon that only happens in non-Hermitian systems.

The rest of the paper is organized as follows. 
We first briefly present the mathematical description of the problem and explain the exponential-growth boundary condition in Sec.~\ref{sec2}. 
Then we build up the PML and apply it to construct the damping eigen-equation in Sec.~\ref{sec3}. 
Finally, we apply our theory to study WGMs and to investigate modes evolution near EPs in Sec.~\ref{sec4}.

\section{Original resonance problem and exponentially growing boundary conditions} \label{sec2}
Because optical devices are fabricated in layered materials, we only consider 2-dimensional (2D) optical resonance problem in this paper. The method could be easily extended to 3-dimensional (3D) resonance problems.

The resonance states are time-steady solutions of Maxwell's equation, in which field components could be decomposed into transverse magnetic (TM) modes and transverse electric (TE) modes. 
We take TM modes as the illustration in the paper, and the formulation could be easily extended for TE modes. 
The stationary field components of TM modes satisfy the Helmholtz-type eigen equation
\begin{equation} \label{eigen_eq}
	\nabla^2 \psi(\boldsymbol{r}) + k^2 n^2(\boldsymbol{r})\psi(\boldsymbol{r}) = 0,
\end{equation}
where $ k^2 $ is the eigenvalue, $ \psi(\boldsymbol{r}) $ is the eigenfunction, and refractive index function $ n(\boldsymbol{r}) $ is regarded as the weight function in the eigen-problem. 
The wave number $ k $ (also stands for resonance frequency $ \omega = ck $ where $ c $ is the light speed constant) is a complex number with real part denoting mode frequency and imaginary part denoting decay rate. 
Field components could be expressed as
\begin{subequations}
	\begin{align}
		&E_r = 0,\quad\quad\quad E_\theta = 0,\quad\quad\quad E_z = \psi,\\
		&B_r = -\dfrac{i}{\omega r}\dfrac{\partial\psi}{\partial\theta},\quad B_\theta = \dfrac{i}{\omega}\dfrac{\partial\psi}{\partial r},\quad B_z = 0.
	\end{align}
\end{subequations}

All field components are associated with the time-dependent factor $ e^{-i\omega t} = e^{-i\Re(k)ct}\cdot e^{\Im(k)ct} $. 
The optical systems considered in the resonance problem are passive cavities, meaning that there is no energy supply once modes are excited. 
Therefore, the energy keeps radiating to infinity and field components are exponentially decaying in time, i.e. $ \Im(k) < 0 $.
Thus we denote $ k = k_r - ik_i $ for some $ k_r,k_i > 0 $, and the quality factor of the modes can be expressed as $ Q = k_r/(2k_i) $.
In fact, when reaching time-steady state, only relative mode intensity remains unchanged.

\begin{figure} [h!]
	\centering
	\includegraphics[width=0.9\columnwidth]{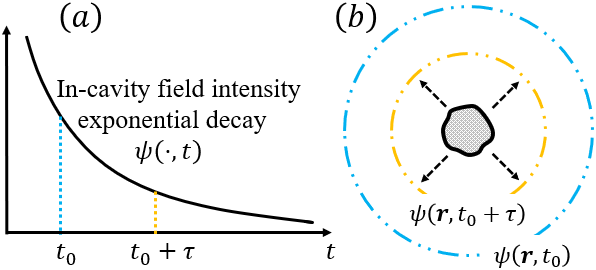}
	\caption{\label{Retardation} Illustration of the retardation effect. (a) field intensity inside cavity goes exponentially decaying in time. (b) spatial orders of wavefront radiated at different times.}
\end{figure}

The exponential-decay in time would result in the field components exponential-growth in space.
This is explained as the retardation effect \cite{retardation}: see Figure~\ref{Retardation},
the wavefront propagating farther away is originated from the cavity at an earlier time, at which time the field amplitude in the cavity is exponentially larger than that of the current moment.
Therefore, in the view of spatial domain at a fixed time, the field component is exponentially growing at far field region. Based on this observation, the radiation boundary condition for the resonance problem is given: as $ r \rightarrow \infty $,
\begin{equation} \label{bc1}
	\psi(r,\theta) \sim \dfrac{F(\theta)}{\sqrt{r}}e^{in_0kr} = \dfrac{F(\theta)}{\sqrt{r}}e^{i n_0 k_r r}\cdot e^{n_0k_ir},
\end{equation}
where we assume the refractive index being constant $ n_0 $ as $ r \rightarrow \infty $. 
In the radiation boundary condition Eq.~(\ref{bc1}), $ F(\theta) $ represents far field pattern; $ e^{i n_0 k_r r} $ is the spatial phase term;
and the exponential growth term $ e^{n_0 k_i r} $ reflects the retardation effect as explain above. 
The denominator $ \sqrt{r} $ is to account for the fact that the 2D cylindrical wavefront propagates in normal direction of the circle with a perimeter $ 2\pi r $, see Figure~\ref{Retardation}(b), meaning that radial component of Poynting vector is proportional to $ 1/r $:
\begin{equation}
	\boldsymbol{S}\cdot\boldsymbol{e}_r = - (E_z e^{-i\omega t} )^* \cdot B_\theta e^{-i\omega t} \sim \dfrac{n_0}{c} \dfrac{\left|F(\theta)\right|^2}{r}.
\end{equation}
For 3D problem, $ \sqrt{r} $ in the boundary condition Eq.~(\ref{bc1}) should be replaced by $ r $, because it is then spherical wavefront propagating in normal direction of the spherical surface with an area $ 4\pi r^2 $.

The radiation boundary condition Eq.~(\ref{bc1}) is similar to but yet different from Sommerfeld radiation condition \cite{sommerfeld}, which is stated as: for some $ k > 0 $,
\begin{equation} \label{bc_sommerfeld}
	\lim\limits_{r\rightarrow\infty} \sqrt{r}\left(\dfrac{\partial \psi}{\partial r} - ik\psi \right) = 0
\end{equation}
holds uniformly in all directions. 
Sommerfeld radiation condition is often applied to ensure there exists a unique solution being physically meaningful for inhomogeneous Helmholtz equation \cite{sommerfeld}.
It describes constant power radiation, whereas Eq.~(\ref{bc1}) describes the radiation from a source which is exponential-decaying in time.
Certainly, the optical resonance problem is an energy-dissipating process and it should subject to the radiation boundary condition in Eq.~(\ref{bc1}).

\section{Construct the damping eigen equation} \label{sec3}
Researchers assimilate the resonance eigen Eq.~(\ref{eigen_eq}) with time-independent Schr\"{o}dinger equation to flourish the study on optical cavities from fruitful results in quantum mechanics \cite{nhqm}. 
The major difference is that standard quantum mechanics requires wavefunction being square-integrable, while the eigenfunctions in the resonance problem as explained in Sec.~\ref{sec2} are exponentially growing at infinity. 
This makes the resonance problems difficult to formulate under the framework of quantum mechanics and even more difficult to solve. 
The perfectly matched layer (PML) method is an ideal technique to transform eigenfunctions from exponential-growth into exponential-decay, and hence the eigenfunctions become square-integrable. 
Accordingly, a non-Hermitian Hamiltonian operator for the optical resonance system can be constructed to calculate its eigenstates.

\subsection{Perfectly matched layer}

\begin{figure} 
	\centering
	\includegraphics[width=0.5\columnwidth]{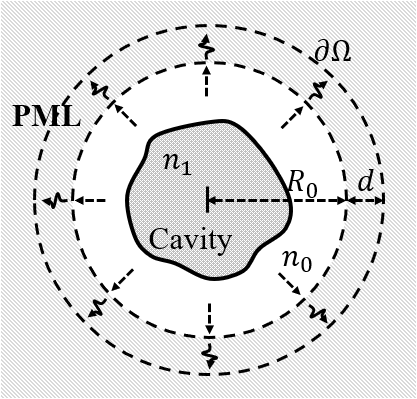}
	\caption{\label{PML} Schematic illustration of perfectly matched layer (PML) in the far field region.}
\end{figure}

As schematically illustrated in Figure~\ref{PML}, the PML method is to introduce from the far field region $ (r > R_0) $ an absorbing layer which is totally free of reflection. 
Because no reflections interfere with inner waves, the eigenfunctions $ \psi(\boldsymbol{r}) $ inside PML $ (r < R_0) $ would preserve as if PML does not exist.
Once penetrating into PML, waves are absorbed when propagating forward, and field amplitude goes exponentially decaying. 

The absorption is introduced by building up a dimensionless damping function $ \tilde{\sigma}(r) $ as
\begin{equation}
	\tilde{\sigma}(r) = \left\{
	\begin{aligned}
		&0 \\
		&\text{increasing}\\
		&\sigma_0
	\end{aligned}
	\right.\quad\quad
	\begin{aligned}
		0 \leq r& < R_0,\\
		R_0 \leq r& < R_0 + d,\\
		r& \geq R_0 + d,
	\end{aligned}
\end{equation}
for some real constant $ \sigma_0 > 0 $. 
In order to be reflectionless, the 2nd order derivative of $ \tilde{\sigma} $ must be continuous, 
however its exact form does not matter.

\subsection{Damping eigen-equation based on complex stretching}
We use $ \tilde{\sigma}(r) $ to build up a complex coordinate $ (\rho,\phi) $ from current real-valued polar coordinate $ (r,\theta) $ through relations:
\begin{equation}
	\rho(r,\theta) = r \left[ 1 + i\tilde{\sigma}(r) \right],\quad\quad \phi(r,\theta) = \theta.
\end{equation}
The complex stretching is expressed by derivative relations of the two coordinate systems:
\begin{equation} \label{complex_stretching}
	\dfrac{\partial}{\partial \rho} = \dfrac{1}{1 + i\sigma(r)}\dfrac{\partial}{\partial r},\quad\quad
	\dfrac{\partial}{\partial \phi} = \dfrac{\partial}{\partial \theta},
\end{equation}
in which, we simplified the expression by introducing the notation:
\begin{equation}
	\sigma(r) \coloneqq \dfrac{d(r\tilde{\sigma})}{dr} = \tilde{\sigma}(r) + r\dfrac{d\tilde{\sigma}(r)}{dr}.
\end{equation}

Replacing the Laplace operator $ \nabla^2_{(r,\theta)} $ in Eq.~(\ref{eigen_eq}) with the complex stretching operator $ \nabla^2_{(\rho,\phi)} $ leads to a damping eigen-equation. To simplify expressions, we also introduce dimensionless variables as
\begin{equation} \label{alphabeta}
	\alpha(r) \coloneqq 1 + i\tilde{\sigma}(r),\quad\quad \beta(r) \coloneqq 1 + i\sigma(r).
\end{equation}
With the relations in Eq.~(\ref{complex_stretching}) and notations Eq.~(\ref{alphabeta}), the damping eigen-equation is expressed in polar coordinates as
\begin{equation}\label{damping_eq}
	\dfrac{1}{\alpha\beta r}\dfrac{\partial}{\partial r}\left( \dfrac{\alpha r}{\beta}\dfrac{\partial \psi}{\partial r} \right) + \dfrac{1}{\alpha^2r^2}\dfrac{\partial^2 \psi}{\partial \theta^2} + k^2n^2(r,\theta)\psi = 0,
\end{equation}
or in Cartesian coordinates as
\begin{align}
	&\dfrac{\partial}{\partial x} \left[ \left( \dfrac{x^2}{\beta^2} + \dfrac{y^2}{\alpha^2} \right) \dfrac{1}{r^2} \dfrac{\partial\psi}{\partial x} + \left( \dfrac{1}{\beta^2} - \dfrac{1}{\alpha^2} \right)\dfrac{xy}{r^2} \dfrac{\partial\psi}{\partial y} \right] \nonumber\\
	+&\dfrac{\partial}{\partial y} \left[ \left( \dfrac{1}{\beta^2} - \dfrac{1}{\alpha^2} \right)\dfrac{xy}{r^2} \dfrac{\partial\psi}{\partial x} + \left( \dfrac{x^2}{\alpha^2} + \dfrac{y^2}{\beta^2} \right) \dfrac{1}{r^2} \dfrac{\partial\psi}{\partial y} \right] \nonumber\\
	+&\dfrac{1}{\alpha\beta^3}\dfrac{d(\alpha\beta)}{dr} \left(\dfrac{x}{r}\dfrac{\partial\psi}{\partial x} + \dfrac{y}{r}\dfrac{\partial\psi}{\partial y}\right) + k^2 n^2(x,y)\psi = 0.
\end{align}
By introducing a 2-by-2 matrix $ A(\boldsymbol{r}) $ as
\begin{equation}
	A(\boldsymbol{r}) \coloneqq \dfrac{1}{r^2}
	\begin{pmatrix}
		\dfrac{x^2}{\beta^2} + \dfrac{y^2}{\alpha^2} & \dfrac{xy}{\beta^2} - \dfrac{xy}{\alpha^2} \\
		\dfrac{xy}{\beta^2} - \dfrac{xy}{\alpha^2} & \dfrac{x^2}{\alpha^2} + \dfrac{y^2}{\beta^2} \\
	\end{pmatrix},
\end{equation}
the damping eigen-equation becomes
\begin{equation} \label{damping_eq_simple}
	\nabla\cdot(A\nabla\psi) + \dfrac{1}{\alpha\beta^3}\nabla(\alpha\beta)\cdot \nabla\psi + k^2 n^2(\boldsymbol{r})\psi = 0.
\end{equation}

With the complex stretching technique transforming the original resonance Eq.~(\ref{eigen_eq}) into the damping eigen-equation Eq.~(\ref{damping_eq_simple}), the technique could further transform the boundary conditions from exponential growth to exponential decay.

Replacing $ (r,\theta) $ with $ (\rho,\phi) $ in the radiation boundary condition Eq.~(\ref{bc1}), it becomes: as $ r \rightarrow \infty $
\begin{align} \label{bc_decay}
	\psi(r, \theta)\sim&\dfrac{F(\phi)}{\sqrt{\rho}} e^{in_0k\rho} \nonumber\\
	=& \dfrac{F(\theta)}{\sqrt{(1 + i\sigma_0)r}} e^{i n_0 (k_r + k_i\sigma_0)r} e^{-n_0 (k_r\sigma_0 - k_i)r}.
\end{align}
This shows that if the constant $ \sigma_0 $ is preset large enough, eigenfunctions of resonance problems are exponentially decaying (hence, square-integrable) after performing the complex stretching.

Therefore, with the damping eigen-Eq.~(\ref{damping_eq_simple}), the optical resonance system is assimilated to a quantum system:
\begin{equation} \label{hamiltonian_eq}
	\hat{\mathcal{H}}\psi = k^2n^2(\boldsymbol{r})\psi,
\end{equation}
in which, the Hamiltonian operator $ \hat{\mathcal{H}} $ is defined as
\begin{equation} \label{hamiltonian}
	\hat{\mathcal{H}} \psi \coloneqq - \nabla\cdot(A\nabla\psi) - \dfrac{1}{\alpha\beta^3}\nabla(\alpha\beta)\cdot \nabla\psi,
\end{equation}
and the eigenfunctions $ \psi $ are subject to the exponential-decay boundary condition Eq.~(\ref{bc_decay}).
The adjoint $ \hat{\mathcal{H}}^\dagger $ of the Hamiltonian operator also can be derived:
\begin{equation}
	\hat{\mathcal{H}}^\dagger \psi = - \nabla\cdot(A^* \nabla\psi) + \nabla\cdot \left( \dfrac{1}{(\alpha\beta^3)^*}\nabla(\alpha\beta)^* \psi \right),
\end{equation}
where asterisk denotes complex conjugate. It's clear to see $ \hat{\mathcal{H}} \neq \hat{\mathcal{H}}^\dagger $, therefore the eigenvalues of $ \hat{\mathcal{H}} $ are complex and the Hamiltonian is non-Hermitian.

This is the reflect of the fact that optical cavities are non-Hermitian systems and hence their Hamiltonian is also non-Hermitian. In the original resonance problem, the Laplace operator in the eigen-Eq.~(\ref{eigen_eq}) is self-adjoint under certain restrictions, whereas the radiation boundary condition Eq.~(\ref{bc1}) carries the non-Hermitian property. After the perfectly matched layer method, the boundary condition Eq.~(\ref{bc_decay}) becomes proper, while the non-Hermitian property transfers to the Hamiltonian $ \hat{\mathcal{H}} $.

\subsection{Matrix form of the damping eigen-equation}
We derive the matrix form of the damping eigen-equation in a cutoff region.
Although PML is built up from far field region and extend to infinity, it is sufficient to cut off PML where field component decays almost to vanished \cite{PML2009}. 
Here we cut off PML at a finite width $ d $, see Figure~\ref{PML}, and restrict the problem in the circular domain $ \Omega $ with radius $ R_0 + d $. 
Because the field component decays to vanished, we apply Dirichlet boundary condition to the outer edge of PML:
\begin{equation} \label{bc_dirichlet}
\psi |_{\partial \Omega} = 0.
\end{equation}

Since eigenfunctions are square-integrable, we look for solutions in Hilbert space, i.e. $ \psi \in H(\Omega) $. 
Here, the notation $ H(\Omega) $ means the Hilbert space of all functions defined on $ \Omega $ with derivatives continuous and subject to Eq.~(\ref{bc_dirichlet}).

For arbitrary wave functions $ \psi_1(\boldsymbol{r}),\psi_2(\boldsymbol{r})$ in the Hilbert space $ H(\Omega) $, we define the bilinear form $ P(\psi_1,\psi_2) $ as the coupling coefficient of the two states:
\begin{align}
	&P(\psi_1,\psi_2) \coloneqq \braket{\psi_1|\hat{\mathcal{H}}|\psi_2}  \nonumber \\
	&= \int_{\Omega} \left[ \nabla \psi_1^* \cdot (A\nabla \psi_2) - \dfrac{\psi_1^*}{\alpha\beta^3}\nabla(\alpha\beta)\cdot \nabla \psi_2 \right] d\boldsymbol{r}.
\end{align}

In the eigen problem Eq.~(\ref{hamiltonian_eq}), $ n^2(\boldsymbol{r}) $ is interpreted as the weight function in the context of Strum-Liouville's problem (although the $ \hat{\mathcal{H}} $ is non-Hermitian), where the inner product for the Hilbert space $ H(\Omega) $ is defined as the bilinear form
\begin{equation}
	Q(\psi_1,\psi_2) \coloneqq \braket{\psi_1|\psi_2} = \int_{\Omega} n^2(\boldsymbol{r}) \psi_1^* \psi_2 d\boldsymbol{r}.
\end{equation}
Specially when $ \psi_1\equiv \psi_2 \equiv \psi $, the inner product $ Q(\psi,\psi) $ represents the total energy stored in the dielectric cavities for the mode $ \ket{\psi} $. 

By performing integration by parts, the weak form of Eq.~(\ref{damping_eq_simple}) can be derived as: for all $ \varphi $ in $ H(\Omega) $,
\begin{equation} \label{weakform}
	P(\varphi,\psi) = k^2 Q(\varphi,\psi).
\end{equation}
Considering varying $ \varphi $ as all $ \varphi_k \in H(\Omega) $ and represent $ \psi $ as linear combinations of all elements in the Hilbert space:
\begin{equation}
	\psi = \sum_{\varphi_j \in H(\Omega)} c_j \varphi_j,
\end{equation}
the weak form Eq.~(\ref{weakform}) is transformed into an algebraic eigen equation as
\begin{equation} \label{matrixeq}
	\widetilde{P}\Psi = k^2 \widetilde{Q}\Psi,
\end{equation}
where the eigenvector is $ \Psi = (\cdots, c_{j}, c_{j+1}, \cdots)^T $ and the matrix entities are
\begin{equation} 
	\widetilde{P}_{ij} = P(\varphi_i, \varphi_j),\quad \quad \widetilde{Q}_{ij} = Q(\varphi_i, \varphi_j),
\end{equation}
for matrix $ \widetilde{P} $ and $ \widetilde{Q} $, respectively. We solve the algebraic eigen Eq.~(\ref{matrixeq}) to calculate the eigen solutions of $ \hat{\mathcal{H}} $.

We remark here that the method presented in this paper is not restricted for single-cavity system, because we do not impose any requirements on the refractive index distributions $ n(\boldsymbol{r}) $. In fact, the presenting method is applicable to any distributions $ n(\boldsymbol{r}) $, including largely deformed cavities, multiple cavities, random media, gradually varied $ n(\boldsymbol{r}) $ etc.

\section{Applications} \label{sec4}
In calculating the eigenstates of $ \hat{\mathcal{H}} $, we set the parameters $ R_0/r_0 = 3 $ and $ d/r_0 = 1 $. The damping function is set as
\begin{equation}
	\tilde{\sigma}(r) = \left\{
	\begin{aligned}
		0& \\
		(r -& R_0)^4
		\end{aligned}
	\right.\quad
	\begin{aligned}
		r& \leq R_0, \\
		R_0 < r& \leq R_0 + d,
	\end{aligned}
\end{equation}
where the region $ r > R_0 + d $ is cut off.

\subsection{Whispering-gallery modes in disk cavities}

\begin{figure} [h!]
	\centering
	\includegraphics[width=\columnwidth]{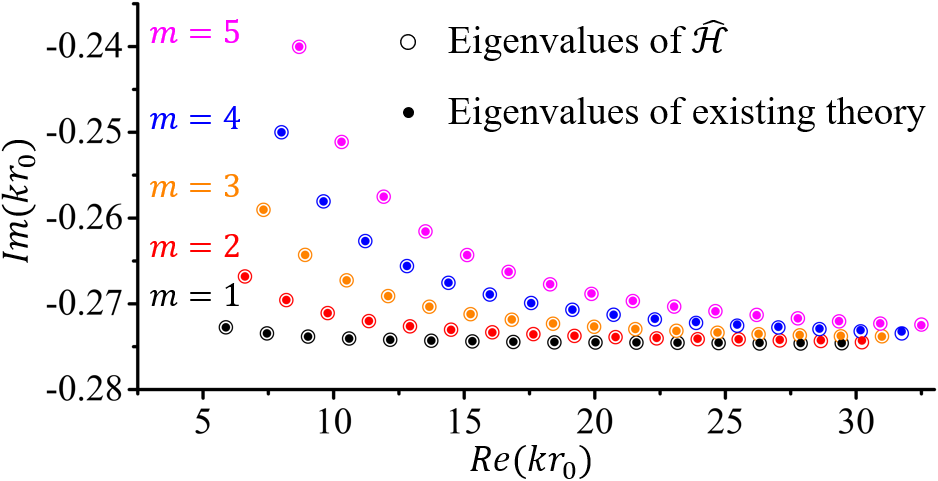}
	\caption{\label{CircularSpectrum} Dimensionless eigenvalues of whispering-gallery modes in the circular cavity for azimuthal order $ m = 1,2,3,4,5 $ and radial order $ l = 4,5,6,\cdots 19 $. Empty spots are eigenvalues of $ \hat{\mathcal{H}} $; filled spots are eigenvalues from solving Eq.~(\ref{eigenveq}).
	}
\end{figure}

In the first calculation, we consider circular-shape cavity with radius $ r_0 = 1 $. The refractive index inside cavity is $ n_1 = 2 $ and outside cavity is $ n_0 = 1 $. We calculate the eigenvalues of $ \hat{\mathcal{H}} $, plotted in Figure~\ref{CircularSpectrum}.

For the circular-shape cavity, eigen-solutions can also be solved analytically and they are called whispering-gallery modes (WGMs) \cite{WGMs1910}. The eigenvalues of WGMs are given by solving the transcendental equation \cite{WGMs2002,perturbation2008}:
\begin{equation} \label{eigenveq}
n_1 \dfrac{J_m' (n_1 k r_0)}{J_m (n_1 k r_0)} = n_0 \dfrac{H_m' (n_0 k r_0)}{H_m (n_0 k r_0)},
\end{equation}
where $ J_m $ and $ H_m $ are Bessel function of $ m $'th order and first-type Hankel function of $ m $'th order, respectively. For each integer $ m $, roots of Eq.~(\ref{eigenveq}) can be found and rearranged in absolute-value ascending order, indexed by integer $ l $. Then each mode could be referred by the mode number $ (m,l) $, where $ m $ is called azimuthal order and $ l $ is called radial order.

By solving Eq.~(\ref{eigenveq}), we find eigenvalues of WGMs for $ m = 1,2,3,4,5 $ and $ l = 4,5,6, \cdots 19 $, also plotted in Figure~\ref{CircularSpectrum}.
The eigenvalues of $ \hat{\mathcal{H}} $ agrees perfectly with the eigenvalues via solving Eq.~(\ref{eigenveq}), and the maximum relative error $ | \Delta k / k | < 1.4 \times 10^{-4} $. The perfectly agreed results validate that the proposed Hamiltonian Eq.~(\ref{hamiltonian}) is very effective for calculating optical modes.

\subsection{Modes evolution near exceptional points in quad-cosine cavities}
In the second calculation, we apply the effective Hamiltonian to quad-cosine cavities to study mode evolution near exceptional points (EPs). 
The phenomenon of EPs happens when the matrix representation of the quantum system is in Jordan form \cite{Eps2012}, meaning that system's algebraic multiplicity is lager than geometric multiplicity. EPs could only happen in non-Hermitian systems, because Hermitian quantum systems in matrix representations are always diagonalizable. 

The quad-cosine cavity is expressed in polar system as
\begin{equation}
	R(\theta) = r_0\left[ 1 + \epsilon \cos(4\theta) \right],
\end{equation}
where cavity radius $ r_0 = 1 $ and $ \epsilon $ is the deformation parameter. 
The refractive index inside the cavity is $ n_1 = 1.6366 $ and outside the cavity is $ n_0 = 1 $. 

By varying the deformation parameter $ \epsilon $, we study mode evolution of the two modes with index $ (24,1) $ and $ (20,2) $. 
Although each mode is associated with clockwise and counter-clockwise (2-fold) degeneracy, we only consider one pair from the two modes, and the behavior of the other pair is similar.

\begin{figure}
	\centering
	\includegraphics[width=\columnwidth]{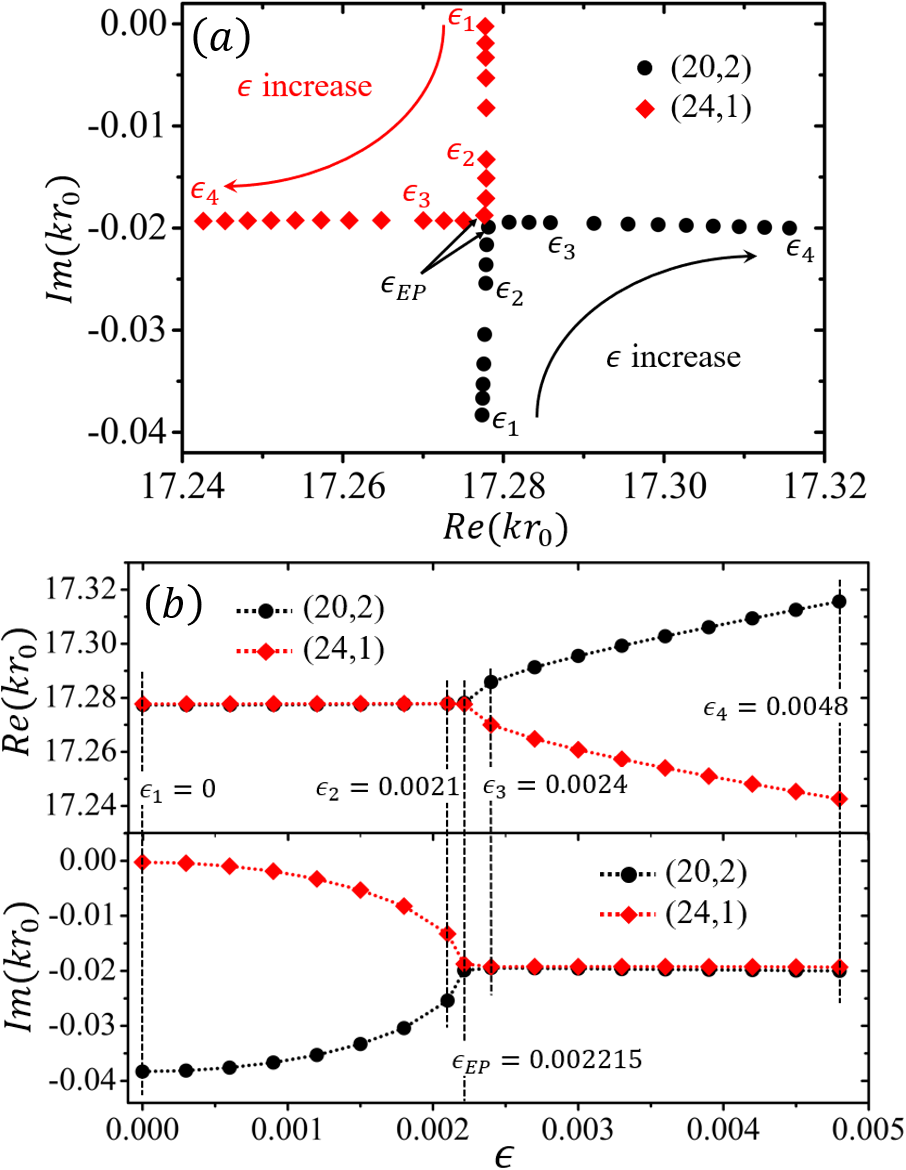}
	\caption{\label{EPeigenvalue} Dimensionless eigenvalues of $ \hat{\mathcal{H}} $ for quad-cosine cavity. (a) Eigenvalues in the complex plane indexed by $ \epsilon $. (b) Real part and imaginary part of eigenvalues for varying $ \epsilon $. Please note that eigenvalues near $ \epsilon_{EP} $ in (a) are more densely plotted than (b).
	}
\end{figure}

\begin{figure*}
	\centering
	\includegraphics[width=2\columnwidth]{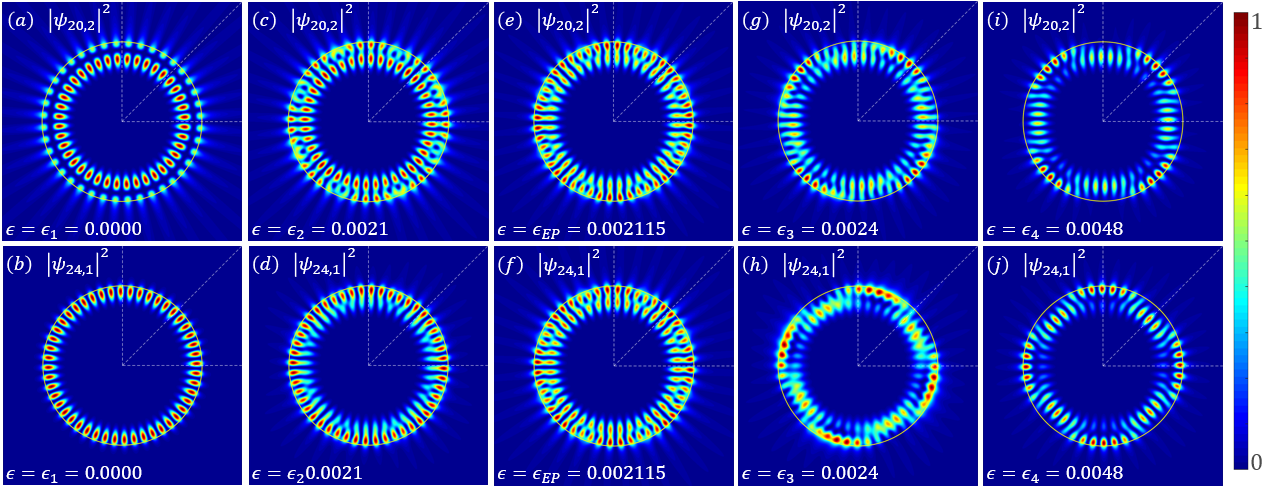}
	\caption{\label{EPdistribution} Mode distributions $ |\psi|^2 $ of $ (20,2) $ and $ (24,1) $: (a) and (b) $ \epsilon = \epsilon_1 = 0 $; (c) and (d) $ \epsilon = \epsilon_{2} = 0.0021 $; (e) and (f) $ \epsilon = \epsilon_{EP} = 0.002215 $; (g) and (h) $ \epsilon = \epsilon_{3} = 0.0024 $; (i) and (j) $ \epsilon = \epsilon_{4} = 0.0048 $. The dash lines in each plot are $ x $-positive, $ y $-positive and $ x = y $ in the first quarter, to help to identify mode distributions.}
\end{figure*}

We calculate the eigenvalues of $ \hat{\mathcal{H}} $ for $ \epsilon $ varying from $ 0 $ to $ 0.0048 $, as shown in Figure~\ref{EPeigenvalue}. We find that $ \epsilon = \epsilon_{EP} = 0.002215 $ is a second-order exceptional point. When $ \epsilon \in [0,\epsilon_{EP}) $, the real parts of eigenvalues are the same and imaginary parts converge as $ \epsilon $ increase. When $ \epsilon $ reaches the exceptional point $ \epsilon_{EP} $, both of the real parts and imaginary parts coalesce. The exceptional point is the turning point. When $ \epsilon $ is larger than $ \epsilon_{EP} $, the imaginary parts keeps unchanged whereas the real part diverge as $ \epsilon $ increase. In Figure~\ref{EPeigenvalue}, we specially mark the deformation parameter in 5 values: $ \epsilon_{1} = 0, \epsilon_{2} = 0.0021, \epsilon_{EP} = 0.002215, \epsilon_{3} = 0.0024 $ and $ \epsilon_{4} = 0.0048 $. We take the 5 samples to investigate the evolution of modes distributions in the following.


Figure~\ref{EPdistribution} shows the evolution of mode distribution of $ (20,2) $ and $ (24,1) $ for $ \epsilon = \epsilon_{1}, \epsilon_{2}, \epsilon_{EP}, \epsilon_{3} $ and $ \epsilon_{4} $. 
In the undeformed cavity $ \epsilon = \epsilon_{1} $, modes are WGMs and they are symmetric and distinctly different (associated with different azimuthal order and radial order). 
When $ \epsilon $ increase, the mode distributions start to become non-symmetric because of the cavity deformation breaks the symmetry.
When $ \epsilon $ is close to $ \epsilon_{EP} $, the two modes start to assimilate to each other.
When $ \epsilon = \epsilon_{EP} $, the two mode distributions become identical, see Figure~\ref{EPdistribution} (e) and (f). 
This is the exceptional point that the eigenstates of the two modes coalesce. 
When $ \epsilon $ continues to increase, the two modes start to couple. 
The two modes are highly coupled when the cavity is largely deformed $ \epsilon = \epsilon_{4} $. 

The successful application of our theory in investigating modes evolution near EPs demonstrates that the proposed Hamiltonian is efficient to reflect the non-Hermitian characteristics in the optical resonance systems.

\section{Conclusion}
In this paper, based on perfectly matched layer we develop a novel method to calculate optical modes in cavities with arbitrary dielectric distributions.
The main mechanism is to introduce at far field region an absorbing layer which is free of reflection. 
We explain the exponential-growth boundary condition as the retardation effect.
With the complex stretching technique, exponentially growing boundary conditions are transformed into exponentially decaying boundary conditions.  
The damping eigen equation is also build up with the complex stretching technique.
We apply our theory to circular cavities calculating the WGMs, and the results perfectly agree with the existing theory which was developed for regularly-shaped cavities. 
Our method is also successfully applied to study the modes evolution near EPs in quad-cosine cavities. 
This indicates the proposed Hamiltonian successfully reflects the non-Hermitian characteristics in optical resonance systems.
Our method does not impose requirements on dielectric distributions, and our method is applicable to resonance systems with arbitrary dielectric distributions, which may facilitate potential studies on resonance in largely-deformed cavities, multi-cavities and random media etc. 
The method imposes the eigenfunctions in a certain Hilbert space, and this helps to formulate the optical resonance problems under the framework of quantum mechanics.


\section*{Acknowledgments}
The authors are grateful to Prof Zhiming Chen of Chinese Academy of Sciences for the fruitful discussions.


%

\end{document}